\documentclass[aps,prl,reprint,twocolumn,superscriptaddress,floatfix,nofootinbib,longbibliography]{revtex4-2}
\usepackage{graphicx,amsmath,amsfonts,amssymb,amsthm,xr}
\usepackage{epsfig,amsmath,amssymb,color,dsfont,upgreek,physics}
\usepackage{mathrsfs}
\usepackage{mathtools}
\usepackage{adjustbox}
\usepackage{bbold}
\usepackage{float}
\usepackage[caption=false]{subfig}

\usepackage[bookmarks=true,colorlinks,linkcolor=OrangeRed,urlcolor=NavyBlue,citecolor=RoyalBlue]{hyperref}
\usepackage[dvipsnames]{xcolor}
\usepackage{orcidlink}

\usepackage{graphicx}
\usepackage{dcolumn}
\usepackage{bm}
\usepackage{color}

\definecolor{mypurple}{rgb}{0.49,0.18,0.56}

\newcommand{\Zt}{\mathbb{Z}_2}
\newcommand{\rr}{\mathbf{r}}

\usetikzlibrary{arrows.meta}
\usepackage{changes}
\usetikzlibrary{quantikz2}
\usepackage{helvet}
\usepackage[tracking=true]{microtype}
\SetTracking{encoding=*}{90} 

\begin{document}

\preprint{}

\title{Observation of glueball excitations and string breaking in a $2+1$D $\mathbb{Z}_2$ lattice gauge theory on a trapped-ion quantum computer}

\author{Kaidi Xu${}^{\orcidlink{0000-0003-2184-0829}}$}
\thanks{These authors contributed equally to this work.}
\affiliation{Max Planck Institute of Quantum Optics, 85748 Garching, Germany}
\affiliation{Department of Physics and Arnold Sommerfeld Center for Theoretical Physics (ASC), Ludwig Maximilian University of Munich, 80333 Munich, Germany}
\affiliation{Munich Center for Quantum Science and Technology (MCQST), 80799 Munich, Germany}

\author{Umberto Borla${}^{\orcidlink{0000-0002-4224-5335}}$}
\thanks{These authors contributed equally to this work.}
\affiliation{Max Planck Institute of Quantum Optics, 85748 Garching, Germany}
\affiliation{Department of Physics and Arnold Sommerfeld Center for Theoretical Physics (ASC), Ludwig Maximilian University of Munich, 80333 Munich, Germany}
\affiliation{Munich Center for Quantum Science and Technology (MCQST), 80799 Munich, Germany}

\author{Kevin Hemery${}^{\orcidlink{0000-0001-7086-593X}}$}
\affiliation{Quantinuum, Leopoldstr.~180, 80804 Munich, Germany}

\author{Rohan Joshi${}^{\orcidlink{0000-0003-1520-7146}}$}

\affiliation{Max Planck Institute of Quantum Optics, 85748 Garching, Germany}
\affiliation{Department of Physics and Arnold Sommerfeld Center for Theoretical Physics (ASC), Ludwig Maximilian University of Munich, 80333 Munich, Germany}
\affiliation{Munich Center for Quantum Science and Technology (MCQST), 80799 Munich, Germany}

\author{Henrik Dreyer${}^{\orcidlink{0000-0002-1480-6406}}$}
\affiliation{Quantinuum, Leopoldstr.~180, 80804 Munich, Germany}

\author{Enrico Rinaldi${}^{\orcidlink{0000-0003-4134-809X}}$}
\affiliation{Quantinuum, Partnership House, Carlisle Place, London SW1P 1BX, UK}
\affiliation{RIKEN Center for Quantum Computing (RQC), RIKEN, Wako, Saitama 351-0198, Japan}
\affiliation{School of Mathematical Sciences, Queen Mary University of London, Mile End Road, London, E1 4NS, UK}

\author{Jad C.~Halimeh${}^{\orcidlink{0000-0002-0659-7990}}$}
\email{jad.halimeh@lmu.de}
\affiliation{Department of Physics and Arnold Sommerfeld Center for Theoretical Physics (ASC), Ludwig Maximilian University of Munich, 80333 Munich, Germany}
\affiliation{Max Planck Institute of Quantum Optics, 85748 Garching, Germany}
\affiliation{Munich Center for Quantum Science and Technology (MCQST), 80799 Munich, Germany}
\affiliation{Department of Physics, College of Science, Kyung Hee University, Seoul 02447, Republic of Korea}

\begin{abstract}
A major goal of the quantum simulation of high-energy physics (HEP) is to probe real-time nonperturbative far-from-equilibrium quantum processes underlying phenomena such as hadronization in quantum chromodynamics (QCD). The quantum simulation of the dynamics of confining strings and glueballs, both essential aspects of quark confinement, in a controllable first-principles way is an important step towards this goal. Here, we realize a $\mathbb{Z}_2$ lattice gauge theory in $2+1$D with a tunable plaquette term on a \texttt{Quantinuum System Model H2} trapped-ion quantum computer. We implement a shallow depth-6 Trotter circuit on a $6 \times 5$ matter-site square lattice utilizing all $56$ available qubits to execute over $1000$ entangling gates. We prepare far-from-equilibrium initial string configurations that we quench across a range of parameters to observe rich dynamical phenomena, such as the formation of gauge-invariant closed-loop excitations reminiscent of glueballs in QCD and multi-order string breaking accompanied by spontaneous matter creation. We further demonstrate experimentally that the system displays genuine $2+1$D dynamics, as evidenced by string snapshots over time that cannot be trivially mapped to $1+1$D physics. Our results demonstrate digital quantum simulations of nonequilibrium dynamics in a higher-dimensional lattice gauge theory and provide an experimentally accessible setting for phenomena related to confinement physics.
\end{abstract}

\date{\today} 
\maketitle

Gauge theories form the theoretical foundation of modern particle physics and describe the fundamental interactions of nature \cite{Weinberg1995QuantumTheoryFields,Weinberg:2004kv,peskin2018introduction}. Among them, quantum chromodynamics (QCD), the gauge theory of the strong interaction, exhibits the striking phenomenon of confinement, whereby quarks and gluons cannot be isolated as free particles but instead form bound states \cite{Wilson1974ConfinementQuarks,Polyakov1977quarkconfinement}. A hallmark consequence of confinement is the formation of color-electric flux tubes that connect charges, giving rise to string-like structures whose dynamics underlies processes such as hadronization in high-energy collisions \cite{Greensite2003theconfinementproblem,greensite2020introductionconfinementproblem,Bali2001QCDforces}.

\begin{figure*}[!htb]
    \includegraphics[width=0.95\linewidth]{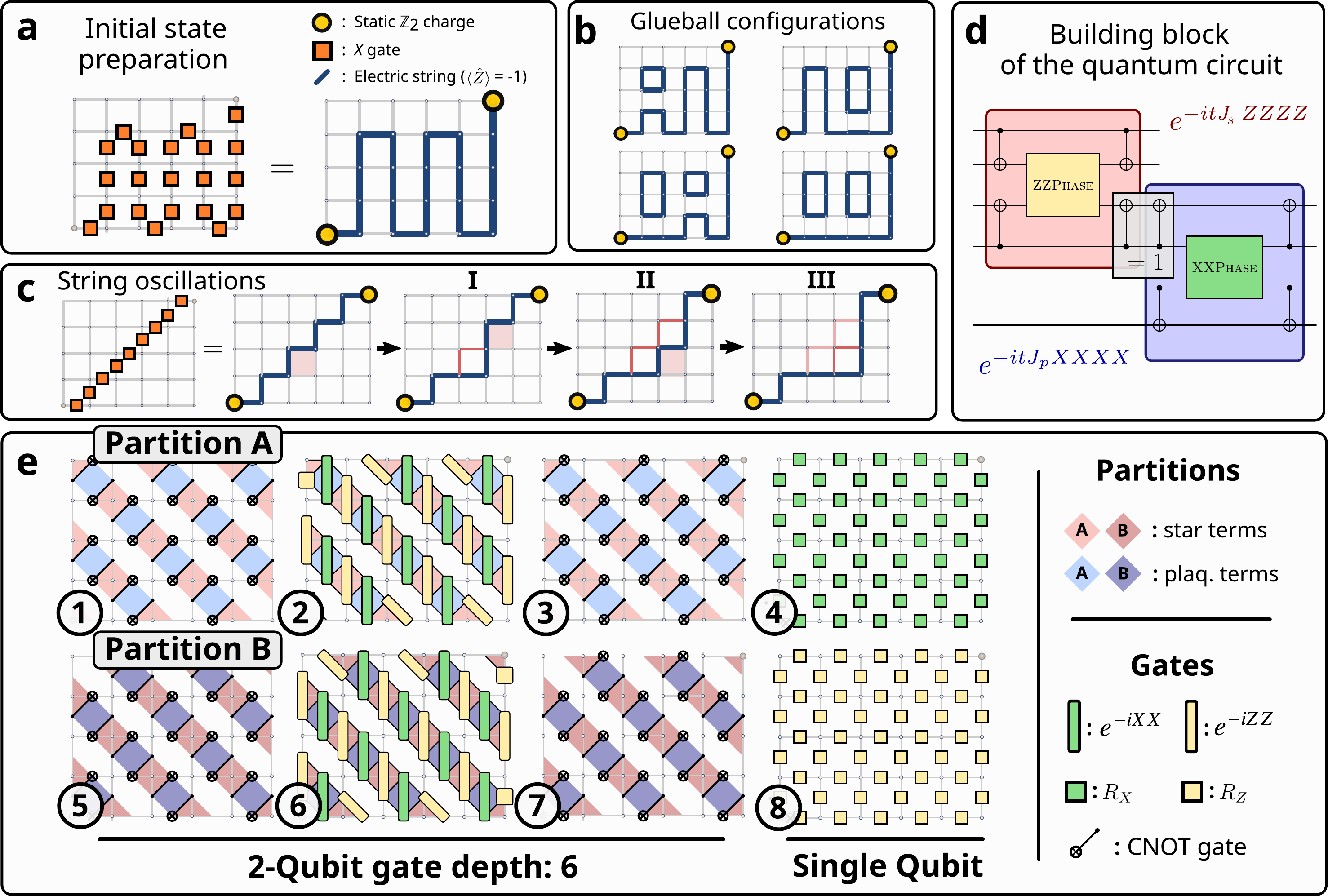}
    \caption{\textbf{Quantum simulation of glueball excitations and string dynamics in a $2+1$D $\mathbb{Z}_2$ lattice gauge theory on a $6 \times 5$ square lattice with open boundary conditions.} \textbf{\textsf{a}} Initial state preparation by acting on the reference state with $X$ gates on the appropriate links, with static $\Zt$ charges encoded in the Hamiltonian. \textbf{\textsf{b}} Examples of ``glueball'' configurations of different number and size (see text). \textbf{\textsf{c}} Illustration of string oscillations in $2+1$D generated by an explicit plaquette operator. \textbf{\textsf{d}} The fundamental building block of our circuit. When acting on an individual partition of the lattice, adjacent CNOT gates contributing to star and plaquette terms cancel off. The $ZZ$ and $XX$ phase gates and the remaining CNOTs can then be moved to the lower layers, dramatically reducing the depth of the circuit. \textbf{\textsf{e}} The quantum circuit protocol for a single Trotter step. Warm-(cool-)colored diamonds correspond to star (plaquette) operators; within each sector, the lighter (darker) shades indicate the partition $p_A$ ($p_B$). At the boundaries, the standard four-body bulk interactions are cleanly truncated, requiring only three-body and two-body phase gates. For each partition, three two-qubit layers are required, followed by single-qubit rotations. 
    }
    \label{fig:quantumcircuit}
\end{figure*}

An especially remarkable prediction of confining gauge theories is the existence of glueballs---bound states composed entirely of gauge fields without any matter constituents \cite{Morningstar1999glueballspectrum,Chen2006glueballspectrum}. In the flux-tube picture of confinement, glueballs can be understood as closed loops of gauge flux, representing collective excitations of the gauge field itself \cite{Isgur1985fluxtubemodel,Meyer2005glueballreggetrajectories}. Despite decades of theoretical and experimental effort, glueballs have remained difficult to identify unambiguously in particle-physics experiments due to mixing with conventional mesonic states \cite{Crede2009theexperimentalstatusofglueballs,Ochs2013thestatusofglueballs}. Their properties have therefore been studied primarily through large-scale Monte Carlo (MC) simulations of lattice gauge theories (LGTs) \cite{Morningstar1999glueballspectrum,Chen2006glueballspectrum,Lucini:2010nv,Gregory:2012hu}, which are lattice formulations \cite{Kogut1975HamiltonianFormulationWilsons,Kogut1979AnIntroductionToLatticeGaugeTheory,Rothe2012LatticeGaugeTheories} of gauge theories originally conceived to study quark confinement \cite{Wilson1974ConfinementQuarks,Wilson1977QuarksStringsLattice}, but which have become powerful tools in condensed matter and quantum many-body physics as well \cite{wen2004quantum,Balents2010SpinLiquidsFrustrated,Savary2016QuantumSpinLiquids,Kleinert1989GaugeFieldsCondensed,Fradkin2013FieldTheoriesCondensed,Smith2017DisorderFreeLocalization,Brenes2018ManyBodyLocalization,Smith2017AbsenceOfErgodicity,Karpov2021DisorderFreeLocalization,Sous2021PhononInducedDisorder,Chakraborty2022DisorderFreeLocalization,Halimeh2022EnhancingDisorderFreeLocalization,Surace2020LatticeGaugeTheories,Lang2022DisorderFreeLocalization,Desaules2023WeakErgodicityBreaking,Desaules2023ProminentQuantumManyBodyScars,Aramthottil2022ScarStates,Tarabunga2023ManyBodyMagic,Desaules2024ergodicitybreaking,Desaules2024MassAssistedLocalDeconfinement,Hudomal2022DrivingQuantumManyBodyScars,Jeyaretnam2025HilbertSpaceFragmentation,Smith2025Nonstabilizerness,Falcao2025nonstabilizerness,Esposito2025magicdiscretelatticegaugetheories,Ciavarella2025GenericHilbertSpaceFragmentation,Ciavarella:2025tdl,Steinegger2025GeometricFragmentationAnomalousThermalization,Ebner2024EntanglementEntropy,Halimeh2023robustquantummany,Iadecola2020QuantumManyBodyScar,Banerjee2021QuantumScarsZeroModes,Biswas2022ScarsFromProtectedZeroModes,Daniel2023BridgingQuantumCriticality,Sau2024sublatticescarsbeyond,Osborne2024QuantumManyBodyScarring,Budde2024QuantumManyBodyScars,Calajo2025QuantumManyBodyScarringNonAbelian,Hartse2025StabilizerScars,cataldi2025disorderfreelocalizationfragmentationnonabelian}.

While such MC calculations have provided important insights into the static spectrum of glueballs, the real-time processes through which gauge-field excitations form and evolve remain largely unexplored \cite{Berges2021qcdthermalization}. This limitation arises because classical computational approaches struggle to access nonequilibrium quantum dynamics in strongly interacting gauge theories. Out of equilibrium, MC techniques suffer from the infamous sign problem \cite{Troyer2005ComputationalComplexityFundamental,Nagata2022FinitedensityLatticeQCD}, while tensor network (TN) methods \cite{Schollwoeck2005DensityMatrixRenormalizationGroup,Schollwock2011DensitymatrixRenormalizationGroup,Orus2014PracticalIntroductionTensorNetworks,Orus2019TensorNetworksComplex,Paeckel2019TimeevolutionMethodsMatrixproduct,Montangero2018IntroductionTensorNetwork,Magnifico2024TensorNetworksLattice} are limited to small system sizes and short evolution times due to the rapid growth of quantum entanglement \cite{Banuls2019TensorNetworksTheir,Rigobello2021EntanglementGeneration$1+1mathrmD$,Xu2025StringBreakingDynamics,cataldi2025realtimestringdynamics21d,cao2026stringbreakingglueballdynamics}.

Quantum simulators offer a promising route to overcome this challenge by enabling controlled implementations of LGTs whose time evolution can be directly observed \cite{Byrnes2006SimulatingLatticeGauge, Dalmonte2016LatticeGaugeTheory, Zohar2015QuantumSimulationsLattice, Aidelsburger:2021mia, Zohar2021QuantumSimulationLattice, 
Barata2022MediumInducedJetBroadening,Klco2022StandardModelPhysics,Barata2023QuantumSimulationInMediumQCDJets,Barata2023RealTimeDynamicsofHyperonSpin, Bauer2023QuantumSimulationHighEnergy, Bauer2023QuantumSimulationFundamental,
DiMeglio2024QuantumComputingHighEnergy, Cheng2024EmergentGaugeTheory, Halimeh2022StabilizingGaugeTheories, Cohen2021QuantumAlgorithmsTransport,Barata2025ProbingCelestialEnergy, Lee2025QuantumComputingEnergy, Turro2024ClassicalQuantumComputing,Halimeh2023ColdatomQuantumSimulators,Bauer2025EfficientUseQuantum,Halimeh2025QuantumSimulationOutofequilibrium}. Recent experiments have begun to realize gauge-theory dynamics on programmable quantum platforms, providing a new avenue to investigate nonperturbative phenomena in real time \cite{Martinez2016RealtimeDynamicsLattice, Klco2018QuantumclassicalComputationSchwinger,Gorg2019RealizationDensitydependentPeierls, Schweizer2019FloquetApproachZ2, Mil2020ScalableRealizationLocal, Yang2020ObservationGaugeInvariance, Wang2022ObservationEmergent$mathbbZ_2$, Su2023ObservationManybodyScarring, Zhou2022ThermalizationDynamicsGauge, Wang2023InterrelatedThermalizationQuantum, Zhang2025ObservationMicroscopicConfinement, Zhu2024ProbingFalseVacuum, Ciavarella2021TrailheadQuantumSimulation, Ciavarella2022PreparationSU3Lattice, Ciavarella2023QuantumSimulationLattice-1, Ciavarella2024QuantumSimulationSU3, 
Gustafson2024PrimitiveQuantumGates, Gustafson2024PrimitiveQuantumGates-1, Lamm2024BlockEncodingsDiscrete, Farrell2023PreparationsQuantumSimulations-1, Farrell2023PreparationsQuantumSimulations, 
Farrell2024ScalableCircuitsPreparing,
Farrell2024QuantumSimulationsHadron, Li2024SequencyHierarchyTruncation, Zemlevskiy2025ScalableQuantumSimulations, Lewis2019QubitModelU1, Atas2021SU2HadronsQuantum, ARahman:2022tkr, Atas2023SimulatingOnedimensionalQuantum, Mendicelli2023RealTimeEvolution, Kavaki2024SquarePlaquettesTriamond, Than2024PhaseDiagramQuantum, Angelides:2023noe, Gyawali2025ObservationDisorderfreeLocalization,  
Mildenberger2025Confinement$$mathbbZ_2$$Lattice, Schuhmacher2025ObservationHadronScattering, Davoudi2025QuantumComputationHadron, Saner2025RealTimeObservationAharonovBohm, Xiang2025RealtimeScatteringFreezeout, Wang2025ObservationInelasticMeson,li2025frameworkquantumsimulationsenergyloss,mark2025observationballisticplasmamemory,froland2025simulatingfullygaugefixedsu2,Hudomal2025ErgodicityBreakingMeetsCriticality,hayata2026onsetthermalizationqdeformedsu2,Cochran2025VisualizingDynamicsCharges, Gonzalez-Cuadra2025ObservationStringBreaking, Crippa2024AnalysisConfinementString, De2024ObservationStringbreakingDynamics, Liu2024StringBreakingMechanism, Alexandrou:2025vaj,Cobos2025RealTimeDynamics2+1D,ilcic2026observationrobustcoherentnonabelian,chen2026thermalizationsu2latticegauge}. In particular, quantum simulation experiments of $2+1$D LGTs now constitute the frontier of the field and can enable the exploration of glueball dynamics (see Fig.~\ref{fig:quantumcircuit}\textbf{a},\textbf{b}), which is not possible in $1+1$D. Despite impressive progress, most existing implementations have lacked an explicit plaquette term \cite{Gyawali2025ObservationDisorderfreeLocalization,Gonzalez-Cuadra2025ObservationStringBreaking,Cobos2025RealTimeDynamics2+1D}, which has been shown to be necessary for genuine $2+1$D string dynamics (see Fig.~\ref{fig:quantumcircuit}\textbf{b},\textbf{c}), itself an essential ingredient for glueball formation \cite{Tian2025RolePlaquetteTerm}. Implementations with an explicit plaquette term have focused on short near-equilibrium strings insufficiently energetic to dynamically generate glueballs \cite{Cochran2025VisualizingDynamicsCharges}.

\begin{figure*}[ht]
    \centering
    \includegraphics[width=0.95\linewidth]{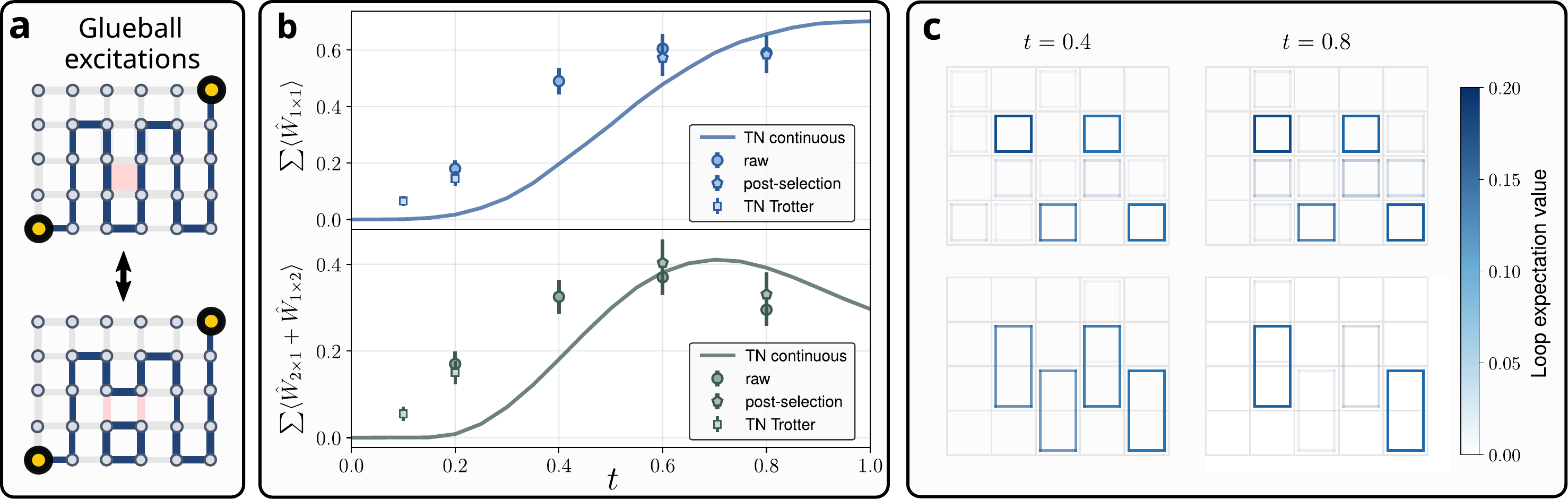}
    \caption{\textbf{Glueball excitations in $2+1$D.} \textbf{\textsf{a}} Example of a ``glueball'' oscillation, where a nonminimal (minimal string refers to the string spanning the minimal length given by the Manhattan distance between two static charges) string transitions into a configuration containing isolated electric loops. \textbf{\textsf{b}} Results for the off-resonance quench of a ``snake'' initial string. The upper and lower panels show the single- and double-plaquette loop expectation values, respectively, covering all the loop configurations reachable with the application of at most two plaquette operators. Here we present the spatially integrated quantities, obtained by summing $\langle \hat{W}_{\Box} \rangle$ over all valid translations of the same loop size across the lattice. \textbf{\textsf{c}} Expectation values of single- and double-plaquette loops at given stroboscopic times, obtained by averaging hardware data over $200$ shots.  We compare the hardware simulation with the continuous TN simulation of the lattice model, and the TN simulation of the Trotterized circuit ($1^\mathrm{st}$ and $2^\mathrm{nd}$ Trotter steps). We also show the results post-selected by leakage detection for the deep circuits ($6^\mathrm{th}$ and $8^\mathrm{th}$ Trotter steps). The parameters are chosen as $J_p=1$, $J_s=7$, $h_{E}=5$, and $h_{M}=0.2$ with a Trotter step $dt=0.1$.}
    \label{fig:glueball}
\end{figure*}

 Here, we consider a $2+1$D $\mathbb{Z}_2$ LGT with dynamical Ising matter, which can be mapped to the toric code with two external fields \cite{Fradkin1979phasediagrams,Trebst2007breakdownofatopologicalphase,Vidal2009lowenergyeffectivetheory,wu2012phasediagram}; see Supplementary Material (SM) \cite{SM}. The model is implemented on a trapped-ion quantum computer through the shallowest known circuit (see Fig.~\ref{fig:quantumcircuit}\textbf{d},\textbf{e}), allowing us to simulate $30$ matter sites ($6 \times 5$) and $49$ gauge links with a tunable plaquette term, surpassing all previous quantum-simulation work in system size. After preparing far-from-equilibrium initial states consisting of electric strings connecting two static $\Zt$ charges, we study their quench dynamics in several parameter regimes. We uncover intriguing dynamics that demonstrate the formation of gauge-invariant closed flux loops resembling glueballs from QCD. We further demonstrate how these glueballs occur in different sizes and robustly persist throughout the all accessible evolution times. By tuning to a resonance, we show different orders of string breaking: a $1^\mathrm{st}$-order resonance breaks the string along a single link with a pair of particles created at the two matter sites to which it connects to screen it; a $2^\mathrm{nd}$-order resonance breaks it along two links, screened by particles created at the two sites to which these links connect. With a tunable plaquette term, we highlight how the uncovered dynamics is genuinely $2+1$D. 

\textbf{Model.---}
The paradigmatic $2+1$D $\mathbb{Z}_2$ LGT with Ising matter integrated out is given by the Hamiltonian 
\begin{align}
    \hat{H}{=}{-}J_s\sum_{s} Q_s\hat{A}_{s} {-} J_{p}\sum_{p} \hat{B}_{p}  {-}  \sum_{i}\big(h_{E} \hat{Z}_{i} {+} h_{M} \hat{X}_{i}\big),
    \label{eq:TC_field}
\end{align}
where the four-body star and plaquette terms $\hat{A}_{s} = \prod_{i\in s} \hat{Z}_{i}$ and $\hat{B}_{p} = \prod_{i \in p} \hat{X}_{i}$ represent the product of Pauli-$Z$ ($X$) operators over the links $i$ emanating from a vertex $s$ and those bounding a plaquette $p$, respectively. The electric field $\hat{Z}_i$ with strength $h_E$ is responsible for the confinement of electric charges, while the minimal coupling $\hat{X}_i$ of strength $h_M$ governs the strength of matter fluctuations, i.e., the hopping and pair creation of dynamical $\Zt$ charges at its endpoint sites. In our convention, links where $\langle\hat Z_i\rangle=-1$ carry electric flux. We use $Q_s=\pm 1$ to denote the absence or presence of static $\Zt$ charges, respectively, i.e., defects which act as a fixed source of electric flux at a site. The fundamental gauge-invariant objects of the model are electric strings, which connect two $\Zt$ charges, and closed electric loops. We focus on their real-time dynamics in the confined regime $h_E \gg h_M, J_p, J_s$, where strings are well defined \cite{Xu2025StringBreakingDynamics}. We work in the electric basis, where $\hat{A}_{s}$ and $\hat{Z}_i$ are diagonal. Initial string configurations connecting two static charges can be expressed as simple product states in the computational basis and are obtained from the initial reference state through the application of $X$ gates; see Fig.~\ref{fig:quantumcircuit}\textbf{a},\textbf{c}. We consider systems with open boundary conditions, enforced by including full four-body plaquettes everywhere on the lattice while restricting $\hat{A}_s$ at the edges and corners to three- and two-body terms, respectively. The two static charges are placed at the bottom-left and top-right corners of the lattice; see Fig.~\ref{fig:quantumcircuit}\textbf{a}.

\textbf{Quantum circuit.---} 
The quantum circuit for each Trotter step is built according to the following partition scheme, aimed at minimizing the two-qubit gate depth of the circuit. As shown in Fig.~\ref{fig:quantumcircuit}\textbf{e}, sets of four links are split into two sublattices, labeled $p_A$ and $p_B$, so that mutually commuting gate operations acting inside one sublattice can be applied in parallel without increasing the two-qubit gate depth. The elementary decomposition of four-qubit terms is illustrated in Fig.~\ref{fig:quantumcircuit}\textbf{d}, 
\begin{equation}
    e^{-itH_\text{4-qubits}}\sim e^{-itJ_s ZZZZ}\,e^{-itJ_p XXXX},
\end{equation}
where a standard entangle-rotation-entangle structure is employed, i.e., one $\mathrm{CNOT}$ layer collects the parity first, then a local phase gate generates the required evolution, and finally the third layer undoes the $\mathrm{CNOT}$s. The noncommuting one-qubit layer is then inserted after executing all the aforementioned operations on each partition. Altogether, an individual Trotter step is organized as 
\begin{equation}
    U(\delta t){\approx}U_{p_B}^{(X)}(\delta t)\,
U_{p_B}^{(Z)}(\delta t)\,U_{1q}^{(II)}\,U_{p_A}^{(X)}(\delta t)\,U_{p_A}^{(Z)}(\delta t)\,U_{1q}^{(I)}.
\end{equation}
The validity of this circuit is tested by comparing it with the exact unitary $U_H=\exp\big({-}it\hat{H}\big)$ as detailed in the SM, where the scaling of the Trotter error is also discussed \cite{SM}. In summary, the total two-qubit gate depth for a single Trotter step is $6$, which, to the best of our knowledge, is the shallowest known toric code circuit. For the $6\times 5$-system, a single Trotter step is composed of $128$ two-qubit gates. At the end of all circuits, the state is destructively measured in the $Z$ basis and the presence of electric field lines, charges, and glueballs is read off from the resulting bitstrings.

\begin{figure}[t!]
    \centering
    \includegraphics[width=0.99\linewidth]{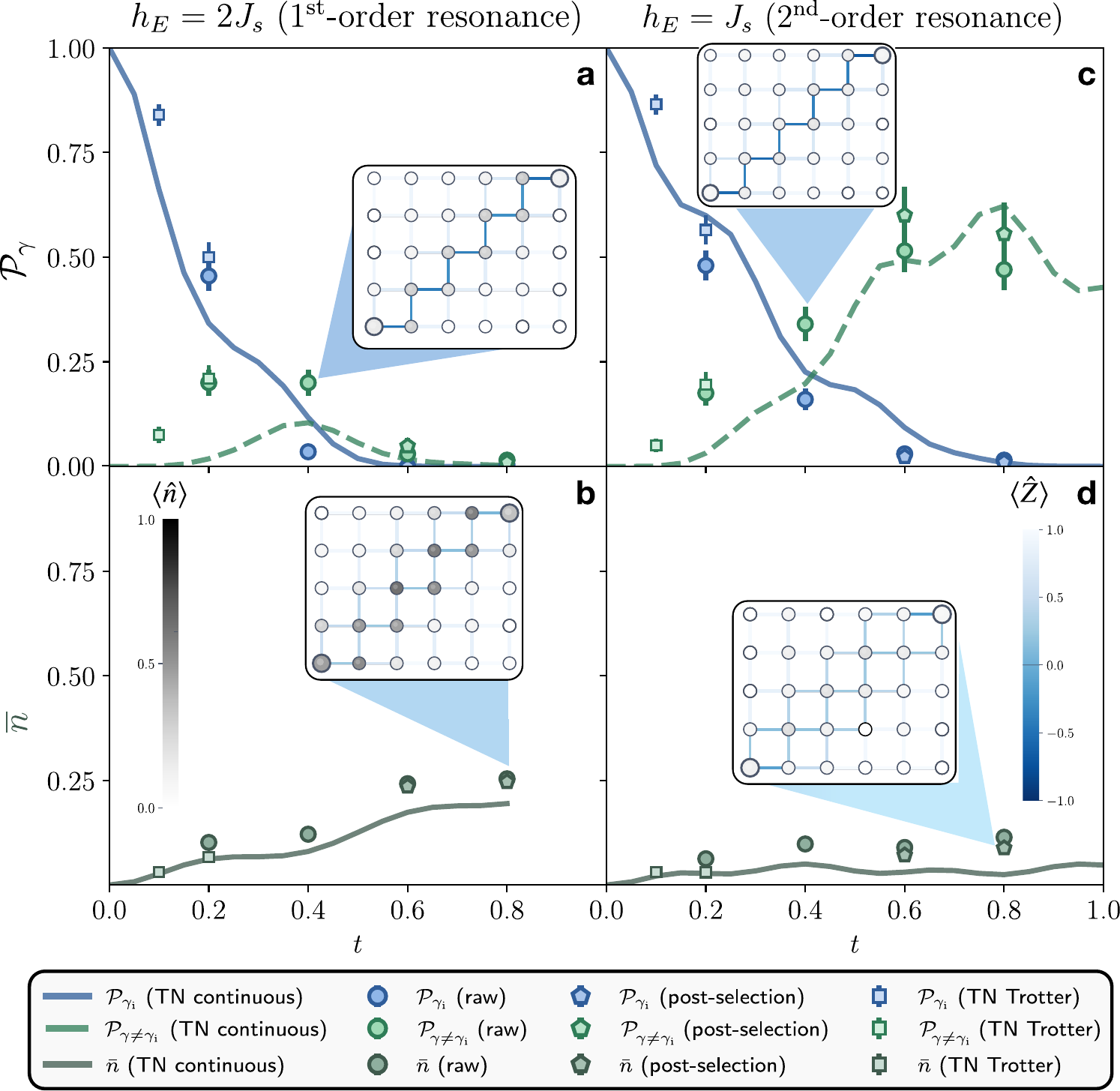}
    \caption{\textbf{String-breaking dynamics in $2+1$D.} The left and right panels show first and second-order resonant string breaking, respectively. The upper panels \textbf{\textsf{a}} and \textbf{\textsf{c}} show the initial string probability $\mathcal{P}_{\gamma_\mathrm{i}}$ and the other minimal-string probabilities $\mathcal{P}_{\gamma \neq \gamma_\mathrm{i}}$, while the lower panels \textbf{\textsf{b}} and \textbf{\textsf{d}} show the particle density $\langle \hat n \rangle$. We compare raw hardware data, leakage--post-selected hardware data, continuous-time TN simulations, and Trotterized TN simulations up to two Trotter steps. The insets show expectation values of matter and gauge observables at given times, obtained by averaging the hardware data over $200$ snapshots. The mass is set to $J_s=2$ and $J_s=4$ for the $1^\mathrm{st}$- and $2^\mathrm{nd}$-order resonance, respectively, with $J_p=1$, $h_E=4$, and $h_{M}=1$.}
    \label{fig:minimal3string}
\end{figure}

\begin{figure*}
    \centering
    \includegraphics[width=0.99\linewidth]{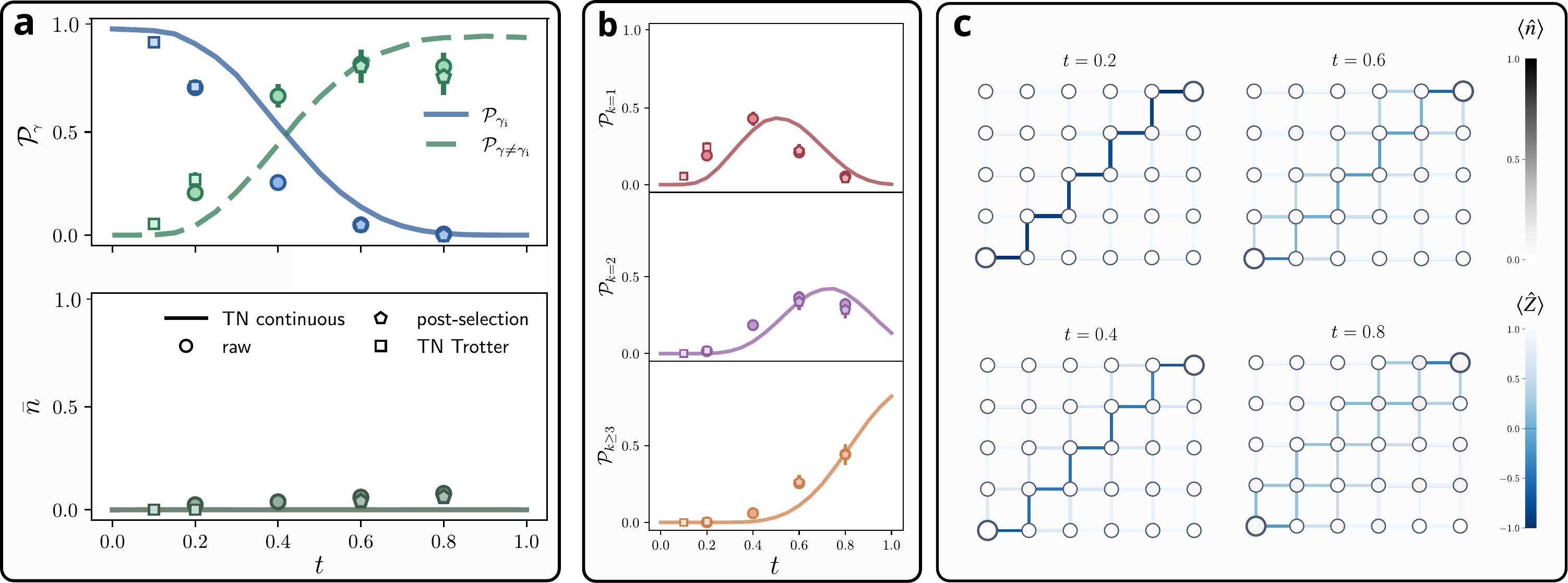}
    \caption{\textbf{Off-resonance quench dynamics of a minimal-length initial string.} \textbf{\textsf{a}} Probabilities of the initial string and other minimal string configurations as well as the average matter density. \textbf{\textsf{b}} The probabilities of the string configurations reached within first-, second-, and third-and-beyond-order plaquette operations. \textbf{\textsf{c}} Expectation values of gauge and matter fields over the whole system, at different stroboscopic times, obtained by averaging hardware data over $200$ snapshots. The parameters are chosen as $J_p=1$, $J_s=7$, $h_{E}=5$, and $h_{M}=0.2$.}
    \label{fig:off-resonance-minimal}
\end{figure*}

\textbf{Glueball dynamics.---}
Quenching long far-from-equilibrium electric flux strings in the confined regime of a $2+1$D $\mathbb{Z}_2$ LGT is expected to lead to rich dynamical signatures that include the formation of electric loops not connected to the original string, resulting in isolated compact pure-gauge configurations bearing resemblance to QCD glueballs \cite{Xu2025StringBreakingDynamics}. Examples of such configurations are shown in Fig.~\ref{fig:quantumcircuit}\textbf{b}. To probe glueball dynamics, we prepare an initial state consisting of an electric ``snake'' string of nonminimal length covering a large portion of the lattice, as shown in Figs.~\ref{fig:quantumcircuit}\textbf{a} and~\ref{fig:glueball}\textbf{a}. After state preparation, the system is evolved under the Trotterized circuit, where each Trotter step is implemented with the aforementioned partition protocol; see Fig.~\ref{fig:quantumcircuit}\textbf{e}. At each stroboscopic time, we evaluate the nonlocal projector from the bitstring outcome from projective measurement onto the computational basis. We define the loop operator as $\hat{W}_{\Box}=\prod_{i\in\Box} \frac{1-\hat{Z}_{i}}{2}$,
where $\Box$ specifies the geometry of the loop, e.g., a single-plaquette ($1\times1$) loop or a double-plaquette ($1\times2$ or $2\times1$) loop.

Figure~\ref{fig:glueball} shows the off-resonant quench dynamics of this ``snake'' initial string for $J_p=1$, $J_s=7$, $h_{E}=5$, and $h_{M}=0.2$. The measured single- and double-plaquette loop probabilities indicate that, at short and intermediate stroboscopic times, the evolution is largely confined to a small set of accessible loop configurations, consistent with coherent glueball-like oscillations of the flux pattern; see Fig.~\ref{fig:glueball}\textbf{b}. The averaged snapshots in Fig.~\ref{fig:glueball}\textbf{c} show the locations and directions on the lattice where the glueballs have proliferated at different evolution times. This spatial inhomogeneity and anisotropy (vertical rather than horizontal double-plaquette glueballs) are directly related to the location and shape of the initial string. Single-plaquette glueballs are most probable at corners of the string where single-order plaquette processes create them. Double-plaquette glueballs are also formed through single-order plaquette processes, but they require a long stretch of the string away from the corner to form. In principle, larger and more exotic glueballs can be formed through first-order plaquette processes by appropriately engineering the shape of the initial string. However, it is essential that the initial string is not of minimal length between the two static charges, as minimal strings cannot lead to glueballs in the confined regime.

We find very good qualitative agreement between TN and hardware results for all probed evolution times in Fig.~\ref{fig:glueball}\textbf{a},\textbf{b} with a moderate amount of shots. After post-selection by leakage detection, a small amplification of the error bar has been observed due to discarded shots, but both the single- and double-plaquette loops are still clearly resolved above the statistical uncertainty. We note that we apply the leakage-detection gadgets only for circuits with 6 and 8 Trotter steps, because the probability of a leakage event increase with the depth of the circuit. Only the first two Trotterized TN time steps, for which we use the matrix product state (MPS) backend in \texttt{Qiskit}, are possible to compute as the bond dimension required for convergence becomes too large at later time steps, indicating that this classical simulation method starts to become too computationally expensive to pursue. The glueball dynamics of the hardware results is faster than the continuous-time TN results, and this is expected from the Trotter error analysis \cite{SM}. Nevertheless, the hardware data show very good qualitative agreement with continuous-time TN simulations, showcasing the hardware capacity to resolve highly nonlocal operator dynamics.

\textbf{String breaking.---}
String breaking occurs at a resonance where the energy gained by flipping one or more links of an electric string, which is equivalent to ``breaking'' it at those links, is compensated by the cost of creating dynamical matter at the broken endpoints. This allows the initial flux tube to convert into shorter strings and mesonic excitations. In a first-order process, a single link is flipped at a cost of $2h_E$, compensated by creating two charges of mass $4J_s$. This fixes the resonance condition to $h_E=2J_s$. Processes where multiple adjacent links are broken, on the other hand, are mediated by higher-order virtual transitions and are only observable over longer timescales. Second-order string breaking, for example, occurs at the resonance condition $h_E=J_s$.

Figure~\ref{fig:minimal3string} compares the dynamics of first- and second-order resonant string breaking obtained from raw hardware data, leakage--post-selected hardware data, continuous-time TN simulations, and Trotterized TN simulations. For the latter, again only the first two Trotter steps converge with respect to bond dimension within our computational power. 

To probe the string-breaking process, we define the string occupation probability $\mathcal{P}_{\gamma} = \langle\prod_{\gamma \in i} \frac{1-\hat{Z}_{i}}{2}\rangle$ for a given string configuration $\gamma$ connecting two static charges. As such, $\mathcal{P}_{\gamma_\mathrm{i}}$ is the probability of the wave function occupying a string along the initial configuration $\gamma_\mathrm{i}$. Similarly, $\mathcal{P}_{\gamma\neq \gamma_\mathrm{i}} = \sum_{\gamma\neq \gamma_{i}}\mathcal{P}_{\gamma}$ is the probability of occupying string configurations $\gamma$ between the two static charges that are away from the initial one. Fixing $J_p=h_M=1$, we see that at first-order resonance ($h_E=2J_s=4$), string breaking occurs at a faster rate than at second-order resonance ($h_E=J_s=4$), as shown by $\mathcal{P}_{\gamma_\mathrm{i}}$ in Fig.~\ref{fig:minimal3string}\textbf{a},\textbf{c}. This leads to faster matter creation and therefore a larger value of it at the same late stroboscopic time $t=0.8$ in the case of the first-order resonance compared to the second-order one, as shown by $\bar{n}$ in Fig.~\ref{fig:minimal3string}\textbf{b},\textbf{d}, which in turn facilitates more prominent $2+1$D string fluctuations in the second-order case, as can be seen in $\mathcal{P}_{\gamma\neq\gamma_\mathrm{i}}$ in Fig.~\ref{fig:minimal3string}\textbf{a},\textbf{c} and the averaged snapshots in Fig.~\ref{fig:minimal3string}\textbf{b},\textbf{d}.

In the case of first-order resonant string breaking, the hardware data already reproduces the main qualitative features of the dynamics with very good quantitative agreement with available Trotterized TN data. Post-selection further filters out some of the spurious matter excitations from bit-flip errors. However, in the case of second-order resonant string breaking, even though qualitative agreement with TN results is still very good, the quantitative discrepancy between hardware and Trotterized TN data is slightly larger. This is expected because second-order string breaking involves only a small amount of charge creation and proceeds through a weaker higher-order process. The physical signal is therefore intrinsically more fragile, and nonphysical background matter generated by bit-flip errors becomes comparatively more important. Post-selecting the states that are not flagged as leaked by the leakage-detection circuit improves agreement with the continuous-time TN data.

\textbf{String oscillations.---} Tuning away from resonance, the string exhibits oscillations in $2+1$D instead of breaking. Figure~\ref{fig:off-resonance-minimal} shows the off-resonant quench dynamics of a minimal initial string, where we set $J_p=1$, $J_s=7$, $h_{E}=5$, and $h_{M}=0.2$. This is marked by the rapid decay (increase) of $\mathcal{P}_{\gamma_\mathrm{i}}$ ($\mathcal{P}_{\gamma\neq\gamma_\mathrm{i}}$) to zero (unity), with matter occupation heavily suppressed throughout the entire probed evolution times; see Fig.~\ref{fig:off-resonance-minimal}\textbf{a}. This indicates that the initial string quickly oscillates over the lattice occupying other minimal string configurations.

One way to make the coherent oscillatory dynamics particularly transparent is to consider sectors that are connected to the initial state by $k$ plaquette flip operations, as illustrated in Fig.~\ref{fig:quantumcircuit}\textbf{c}: Let $\Omega_{k}$ denote the set of all possible string paths corresponding to a $k^\mathrm{th}$-order deformation (such as paths generated by $k$ plaquette flips from the initial configuration). The total probability of finding the string in the $k^\mathrm{th}$-order sector is then given by the sum of the individual occupation probabilities $\mathcal{P}_{k} = \sum_{\gamma \in \Omega_k} \mathcal{P}_{\gamma}$. As shown in Fig.~\ref{fig:off-resonance-minimal}\textbf{b}, the measured sector weights display a clear hierarchy: $\mathcal{P}_{k=1}$ rises first and reaches the largest amplitude, $\mathcal{P}_{k=2}$ builds up over longer timescales with a smaller but still substantial weight, and $\mathcal{P}_{k\ge3} = \sum_{k\ge3} \mathcal{P}_{k}$ appears only later and remains the weakest contribution. The dynamics therefore exhibits an ordered progression in both timescale and magnitude, with lower-order string rearrangements dominating the early-time evolution and higher-order deformations entering only gradually.

The averaged snapshots in Fig.~\ref{fig:off-resonance-minimal}\textbf{c} confirm that this evolution is best understood as coherent string oscillations within a manifold of nearby configurations, rather than rapid delocalization or strong matter-producing dynamics. Consistently, the average matter density remains small throughout the evolution, showing that the quench primarily reshuffles electric flux, and one can also infer from the previous data that the charge creation mainly comes from the bit-flip error. The limited size of the accessible subspace is an important practical advantage, since it allows the dominant features of the evolving state to be resolved directly through the sector probabilities and reconstructed snapshots.

As shown in Fig.~\ref{fig:off-resonance-minimal}\textbf{a},\textbf{b}, the raw and post-selected hardware data not only show very good quantitative agreement between them but also very good qualitative agreement with the continuous-time TN data and very good quantitative agreement with the accessible Trotterized TN data.

\textbf{Conclusion and outlook.---}
In this work, we have realized a $2+1$D $\mathbb{Z}_2$ lattice gauge theory with a tunable plaquette term on a trapped-ion quantum computer and directly observed its real-time dynamics following far-from-equilibrium quenches. Enabled by a highly efficient, shallow circuit implementation that operates at unprecedented system size and gate depth, we access a regime of gauge dynamics beyond previous experiments. We report dynamical signatures of gauge-field bound states in the form of coherent closed flux loops, consistent with excitations reminiscent of QCD glueballs, alongside controlled first- and second-order string breaking processes. By engineering the plaquette interaction, we demonstrate genuinely $2+1$D gauge-field dynamics that cannot be reduced to $1+1$D physics. Our results establish quantum computers as a platform for observing the emergence of composite excitations in strongly interacting gauge theories in real time and open a route toward first-principles studies of hadronization and the formation of bound states in quantum field theories. 

\medskip

\footnotesize

\textit{Note.---}
In the same \texttt{arXiv} listing as this paper, a parallel submission \cite{Joshi2026ObservationOfGenuine$2+1$DStringDynamics} involving some of the current authors will appear, where \texttt{Quantinuum System Model H2} is used to observe genuinely $2+1$D string breaking dynamics in a U$(1)$ LGT on a $5\times4$ square lattice with a tunable plaquette term. While completing this work, we became aware of independent and complementary research by M.~John \textit{et al.}~on the dynamics of non-abelian strings, including gluonic excitations, on a qudit quantum processor.

\medskip

\textbf{Acknowledgments.---}
We thank Michael Foss-Feig, Yuta Kikuchi, and Yizhuo Tian for fruitful discussions and comments on our draft.
K.X., U.B., R.J., and J.C.H.~acknowledge funding by the Max Planck Society, the Deutsche Forschungsgemeinschaft (DFG, German Research Foundation) under Germany’s Excellence Strategy – EXC-2111 – 390814868, and the European Research Council (ERC) under the European Union’s Horizon Europe research and innovation program (Grant Agreement No.~101165667)—ERC Starting Grant QuSiGauge. Views and opinions expressed are, however, those of the author(s) only and do not necessarily reflect those of the European Union or the European Research Council Executive Agency. Neither the European Union nor the granting authority can be held responsible for them. All experiments were run on \texttt{Quantinuum H2-2} quantum computer, powered by \texttt{Honeywell}. This work is part of the Quantum Computing for High-Energy Physics (QC4HEP) working group.
\bibliography{biblio}
\normalsize

\clearpage
\pagebreak
\setcounter{equation}{0}
\setcounter{figure}{0}
\setcounter{table}{0}
\setcounter{page}{1}
\setcounter{section}{0}
\makeatletter
\renewcommand{\theequation}{S\arabic{equation}}
\renewcommand{\thefigure}{S\arabic{figure}}
\renewcommand{\thesection}{S\Roman{section}}
\renewcommand{\thepage}{\arabic{page}}
\renewcommand{\thetable}{S\arabic{table}}
\vspace{0cm}
\normalsize
\onecolumngrid
\begin{center}
    \textbf{\large Supplemental Online Material for \\``Observation of glueball excitations and string breaking in a $2+1$D $\mathbb{Z}_2$ lattice gauge theory on a trapped-ion quantum computer''}\\[5pt]
    \vspace{0.1cm}
    \begin{quote}
    {\small In this Supplemental Material, we detail the model derivation and the numerical techniques employed in this study.}\\[10pt]
    \end{quote}
\end{center}

\section{Gauge theory formulation}
The model discussed in the main text can be exactly mapped to a $\Zt$ lattice gauge theory coupled to Ising matter, described by the Hamiltonian
\begin{equation}
H = -J_s\sum_s \hat\tau^x_s -J_p\sum_{p} \hat B_{p} - h_z \sum_{\rr,\eta} \hat \tau^z_\rr \hat \sigma^z_{\rr,\eta} \hat \tau^z_{\rr+\eta} - h_x \sum_{\rr,\eta} \hat \sigma^x_{\rr,\eta}.
\label{eq:2DFS}
\end{equation}
Here, a new set of Pauli matrices $\hat\tau$ represents Ising matter fields on the sites of the lattice, while the $\hat\sigma$ matrices describe $\Zt$ gauge fields on the links. To obtain \eqref{eq:TC_field}, the Ising matter fields $\hat\tau$ are integrated out by resolving the Gauss law 
\begin{equation}
    \hat \tau^x_s=Q_s \hat A_{s},
\end{equation}
which relates the electric flux going out of a vertex to the $\Zt$ parity on the same vertex, and by fixing the ``unitary'' gauge $\hat \tau^z=+1$. In our specific setup, the background charges are chosen so that $Q_s=+1$ everywhere except for the endpoints of the initial string states. 

\section{Cost}
For all the experiments described in the main text, the circuit corresponding to an individual Trotter step is iterated $n$ to times reach different stroboscopic times $t_n$. The largest number of Trotter steps that we choose is $8$, and we only perform destructive measurements at even times $t_n$ with $n=2,4,6,8$. The costs for different numbers of Trotter steps are listed in Tab.~\ref{cost}. Since we always use the same protocol, only changing the initial state and the parameters in the Hamiltonian, the total cost for the $4$ sets of parameters considered in the main text is the same. For $t_2$ and $t_4$ no leakage detection is needed, and hence the two-qubit gates depth is just $6n$, $6$ being the two-qubit depth of a single Trotter step as explained in the main text. For the deep circuits needed to reach $t_6$ and $t_8$ we use the leakage detection gadget shown in Fig.~\ref{fig:LD}, which increases the gate count significantly. 

\begin{table}[h]
\begin{tabular}{|l|l|l|l|l|l|l|l|l|}
\hline
Trotter Step & Number of Qubits & Leakage Detection & 1qb Gates & 2qb Gates & Depth 2qb & Barrier & Reset & Measurement \\ \hline
2            & 49               & No                     & 493       & 128       & 12        & 0       & 0     & 49          \\ \hline
4            & 49               & No                     & 939       & 256       & 24        & 0       & 0     & 49          \\ \hline
6            & 56               & Yes                     & 1728      & 866       & 49        & 147     & 49    & 98          \\ \hline
8            & 56               & Yes                     & 2174      & 1122      & 61        & 147     & 49    & 98          \\ \hline
\end{tabular}
\caption{The total costs for the experiments in the main text. }
\label{cost}
\end{table}

\section{Trotter error analysis and scaling}

We analyze here how the Trotter error caused by a finite time step $dt$ affects the accuracy of our hardware runs. This can be done by comparing the full unitary dynamics for a single timestep  $U_{\text{full}}(dt) = e^{-i H dt}$ to the actual hardware implementation where two noncommuting unitaries, corresponding to four-body and single-body terms respectively, are applied one after the other. In particular, the electric field does not commute with the plaquette terms, and the matter fluctuations do not commute with the star terms. To understand the scaling of Trotter with respect to the time step and system size, we first apply the Baker--Campbell--Hausdorff formula,

\begin{equation}
\begin{aligned}
& \exp\big[-i(-J_s \sum_s \hat{A}_s - J_p \sum_p \hat{B}_p)\, dt\big]\,
  \exp\big[-i(-h_M \sum_i \hat{X}_i - h_E \sum_i \hat{Z}_i)\, dt\big] \\
& \quad = \Big(\exp\big[-i(-J_s \sum_{s/2} \hat{A}_s - J_p \sum_{p/2} \hat{B}_p)\, dt\big]\Big)^2\,
  \exp\big[-i(-h_M \sum_i \hat{X}_i - h_E \sum_i \hat{Z}_i)\, dt\big] \\
& \quad = \exp\Big[
-idt\Big(-J_s \sum_s \hat{A}_s - J_p \sum_p \hat{B}_p - h_M \sum_l \hat{X}_l - h_E \sum_i \hat{Z}_i\Big) \\
& \qquad\qquad
- \frac{h_M J_s dt^2}{2}\Big[\sum_s \hat{A}_s, \sum_i \hat{X}_i\Big]
- \frac{h_E J_p dt^2}{2}\Big[\sum_p \hat{B}_p, \sum_i \hat{Z}_l\Big] \\
& \qquad\qquad
+ \frac{i h_M J_s^2 dt^3}{12}\Big[\sum_s \hat{A}_s,\Big[\sum_s \hat{A}_s, \sum_i \hat{X}_i\Big]\Big] \\
& \qquad\qquad
+ \frac{i h_M^2 J_s dt^3}{12}\Big[\Big[\sum_s \hat{A}_s, \sum_i \hat{X}_i\Big], \sum_i \hat{X}_i\Big]
+ \dots \Big] \\
& \quad = \exp\Big[
-idt\Big(-J_s \sum_s \hat{A}_s - J_p \sum_p \hat{B}_p - h_M \sum_i \hat{X}_i - h_E \sum_i \hat{Z}_i\Big) \\
& \qquad\qquad
- i dt^2 J_s h_M \sum_s \Big(
2\hat{Y}_{s_0}\hat{Z}_{s_1}\hat{Z}_{s_2}\hat{Z}_{s_3}
+ 2\hat{Z}_{s_0}\hat{Y}_{s_1}\hat{Z}_{s_2}\hat{Z}_{s_3}
+ \dots \Big) \\
& \qquad\qquad
+ i dt^2 J_p h_E \sum_p \Big(
2\hat{Y}_{p_0}\hat{X}_{p_1}\hat{X}_{p_2}\hat{X}_{p_3}
+ 2\hat{X}_{p_0}\hat{Y}_{p_1}\hat{X}_{p_2}\hat{X}_{p_3}
+ \dots \Big) \\
& \qquad\qquad
+ O(dt^3)\Big],
\end{aligned}
\end{equation}
to obtain a unitary operator whose exponent encodes the complete dynamics plus corrections which are quadratic in the timestep and linear in the model coefficients. Since our experiments operate in the confining regime $h_E \gg h_M, J_p, J_s$, the last term $\propto J_p h_E$ is the most relevant source of errors. These, up to a configuration-dependent sign, are plaquette flips. As observed in our experiments, transitions to string configurations induced by the application of plaquette operators indeed appear earlier than in the continuous TN simulations.

While several partitions of the Hamiltonian into commuting terms are possible, we have chosen to decompose the terms into two sub-lattices as this results in short-depth circuits, while making use of the four gate zones of the device in parallel.
\begin{figure}[htbp]
    \centering
    \includegraphics[width=0.8\linewidth]{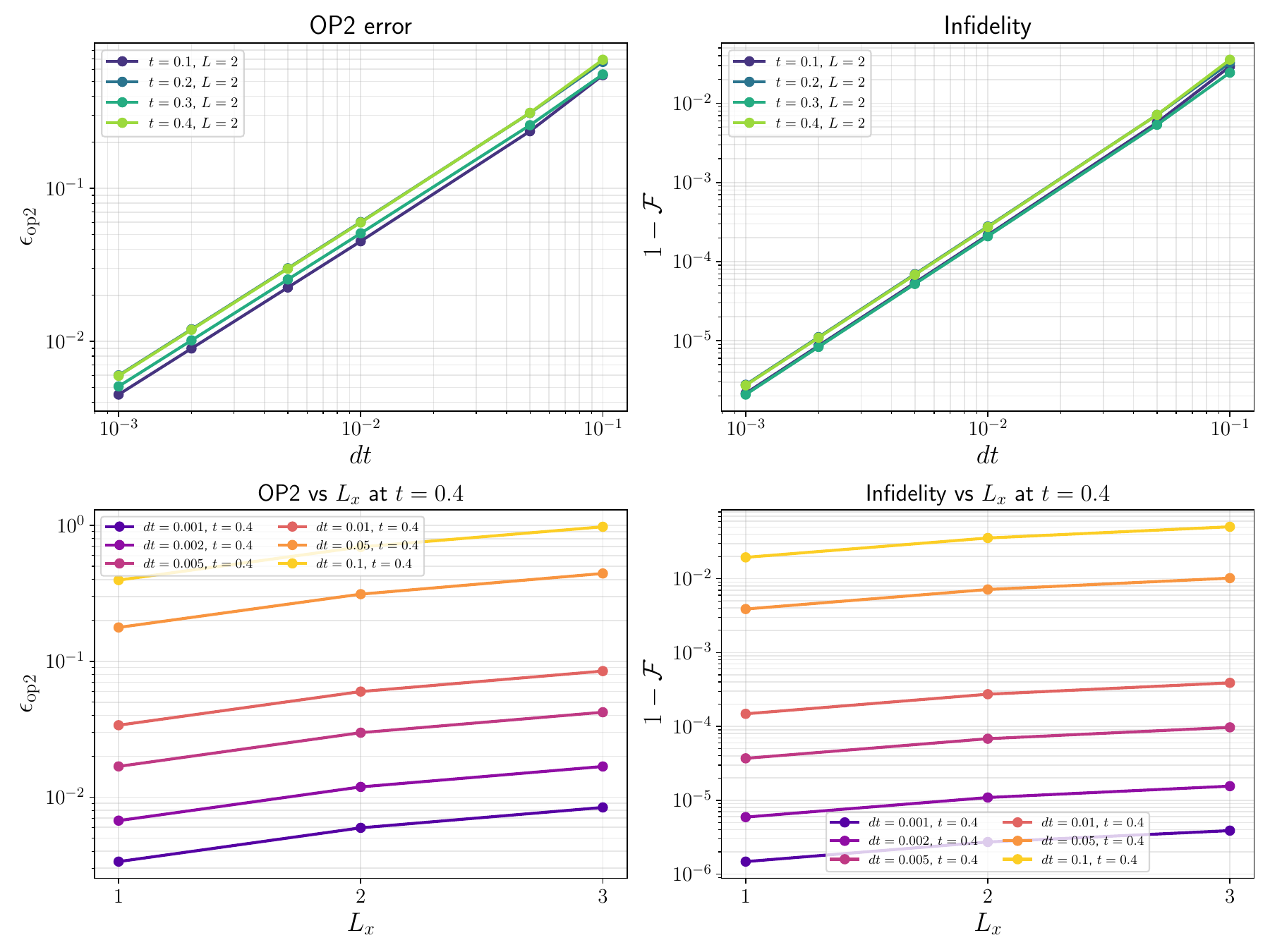}
    \caption{The scaling of Trotter error with respect to the exact unitary. The upper panel shows the log-log scaling of different sizes of Trotter step for $L_x=2$, $L_y=1$ lattice, while the lower panel shows the Trotter error accumulated at t=0.4 for different $L_x$ with $L_y=1$ fixed with $J_p=1$, $J_s=7$, $h_E=5$, $h_M=0.1$.}
    \label{fig:unitary}
\end{figure}
To quantify the difference between the exact and Trotter unitaries, we define the following metrics. The error of the operator $2$-norm is defined as,

\begin{equation}
    \epsilon_{\rm op2}=\left|U_{\rm ref}-e^{-i\phi}U\right|_2, 
\end{equation}
and the global phase $\phi$ is chosen to maximize the overlap. The processing infidelity takes the form

\begin{equation}
    1- \mathcal{F}=1-\frac{\left|\mathrm{Tr}\!\left(U_{\mathrm{ref}}^\dagger U\right)\right|}{d},
\end{equation}
where $d$ is the Hilbert space dimension. As one can see from Fig.~$\ref{fig:unitary}$, as $dt\rightarrow0$, our partition gives the exact unitary and the total accumulated Trotter error in terms of the above two metrics fluctuates around some fixed value (one can see the $t=0.2$ and $t=0.4$ curve almost overlap with each other). For the time step $dt=0.1$ we have taken in the experiments, the total infidelity accumulates $0.024$ after 4 applications. One can also see that both operator 2 error and infidelity doesn't show strong sensitivity against system size from the lower panel of Fig.~\ref{fig:unitary}.

While the fidelity measures how closely the Trotter unitary approximates the continuous-time evolution, our primary concern is the error in the relevant observables, which in this work are the projectors onto the strings. Unfortunately, although the fidelity bounds the error for all observables, it does not provide a straightforward prescription for what level of error is acceptable for our purposes. An excessively small Trotter step can lead to larger deviations from the desired results due to more costly circuits increasing the effect of hardware noise. Classical simulations of the circuits show acceptable agreement with the continuous-time evolution (see Figs. \ref{fig:glueball} and \ref{fig:off-resonance-minimal} in the main text), while ensuring that their cost remains within the regime in which the H2 device operates faithfully. In practice our Trotter step choice ensures good qualitative agreement of the hardware results but introduces a slight time scale shift as explained above.

\section{Observables from projective measurements}

The probability of each string configuration is computed by applying a local projector onto the relevant link qubits. For a given string configuration $\gamma$, the local string projector can be defined as,
\begin{equation}
    \hat{P}_{\gamma} = \prod_{\ell \in \gamma} \hat{P}_\ell^{(1)},
\end{equation}
where the product runs over all link qubits $\ell$ belonging to the configuration $\gamma$, and $\hat{P}_\ell^{(1)} = \ket{1}\bra{1}_\ell = (1-\hat{Z}_l)/2$ projects link $\ell$ onto the state $\ket{1}$. The string probability is then evaluated as the expectation value

\begin{equation}
    p_{\gamma}(t) = \frac{1}{N_{\mathrm{shots}}} \sum_{i=1}^{N_{\mathrm{shots}}} \prod_{\ell \in \gamma} b_\ell^{(i)},
\end{equation}
where $N_{\mathrm{shots}}$ is the total number of measurement shots and $b_{\ell}^{(i)} \in \{0,1\}$ is the eigenvalue of $\hat{P}_{\ell}^{i}$ for the $i^{th}$ shot. This quantity measures the probability that all link qubits along the string configuration $\gamma$ are simultaneously in state $\ket{1}$, irrespective of the state of the remaining qubits not belonging to $\gamma$. For all the simulations, we set $N_{\text{shots}} = 200$.

\section{Details of hardware experiments}
Our experiments use a \texttt{Quantinuum System Model H2} trapped-ion quantum computer \cite{quantinuum-h2-2}, which provides 56 physical qubits with any-to-any connectivity and support for up to four parallel two-qubit operations. 
The processor is based on a QCCD architecture \cite{wineland1998experimental, kielpinski_architecture_2002} with eight gate zones, where qubits encoded in the hyperfine states of $^{171}\mathrm{Yb}^+$ ions are transported between interaction regions and manipulated with laser-driven gates. Four gate zones enable parallel two-qubit gate operations. 
The native gate set includes single-qubit rotations and a parameterized-angle $ZZ$ gate. 

For \texttt{Quantinuum H2-2}, the benchmarked gate infidelities are $2.8\times10^{-5}$ ($\pm 3.6\times10^{-6}$) for single-qubit gates, $8.4\times10^{-4}$ ($\pm 4.8\times10^{-5}$) for two-qubit gates, and $6.7 \times10^{-4}$ ($\pm 8.7\times10^{-5}$) for preparation and measurement in the $|0\rangle$ state.
The memory error per depth-1 circuit time is $1.2\times10^{-4}$ ($\pm 2.0\times10^{-5}$), while the measurement cross-talk error is $2.2\times10^{-5}$ ($\pm 5.3\times10^{-7}$).
These reported infidelities are averaged across all operational zones and are available on GitHub together with the corresponding randomized benchmarking data\cite{quantinuum-performance}.

\begin{figure}[htbp]
    \centering
    \includegraphics[width=0.95\linewidth]{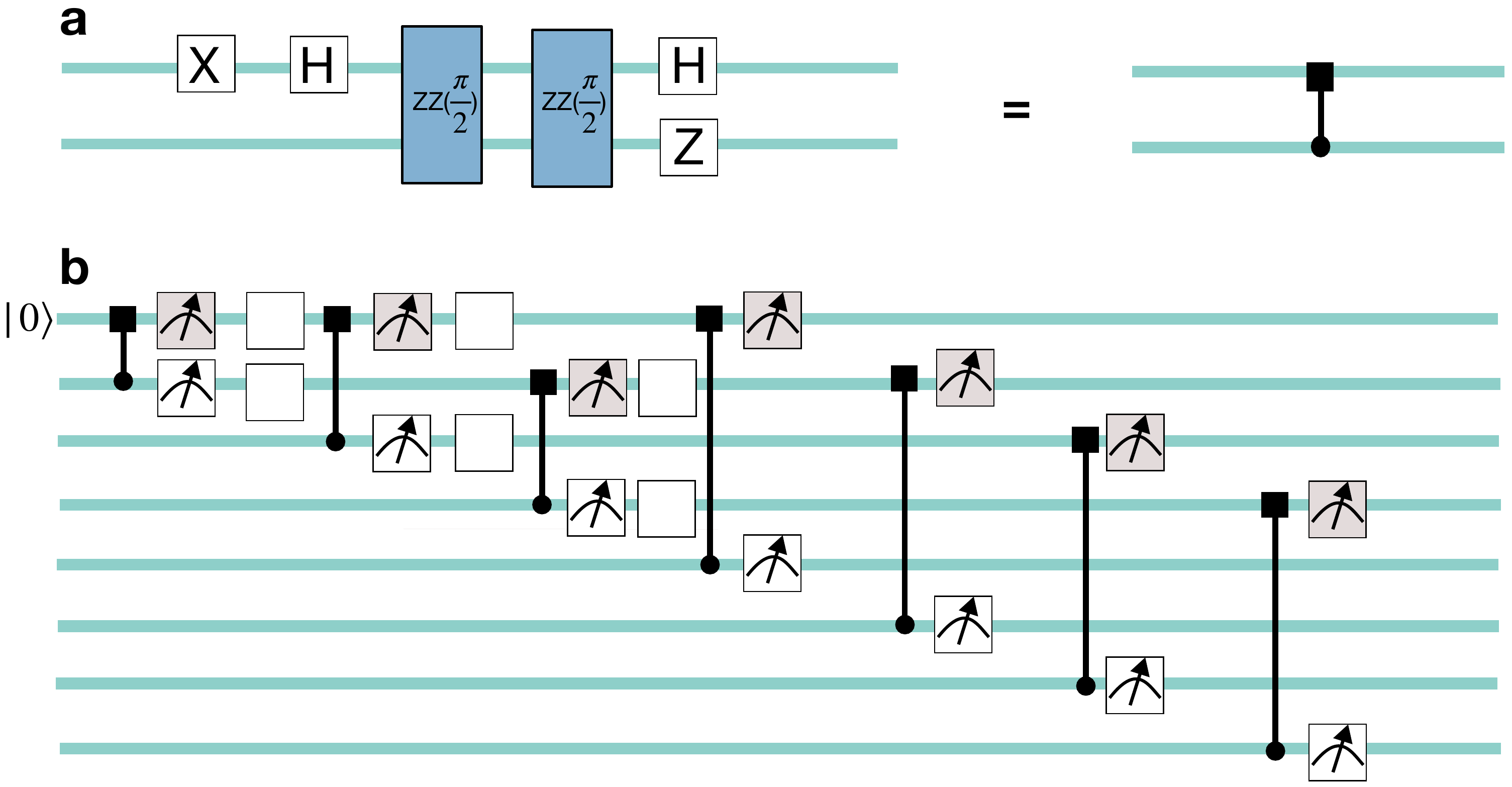}
    \caption{Leakage detection circuit with $a\sim \log(N)$ depth. (a) The gadget. The upper qubit is used to detect whether the lower one is leaked and is initialized as $|0\rangle$. If the lower qubit is leaked, the ancilla state will be $|1\rangle$. (b) An example leakage detection circuit with ancilla reuse. The measurement onto the gadget registers and data registers are distinguished by the colors grey and white, respectively. The white squares represent the reset operation. 
    }
    \label{fig:LD}
\end{figure}

Some amount of error is accounted for by the leakage out of the qubit subspace into other internal states of the $^{171}\mathrm{Yb}^+$ ion. Such leakage events can be detected using the fact that two-qubit gates acting on at least one leaked qubit are effectively deleted from the circuit. A gadget to detect leakage events is shown in Fig.~\ref{fig:LD}. A given data qubit is coupled to an ancilla using two subsequent $\exp(-i \pi/4 ZZ)$ gates, which are logically equivalent to a Pauli $ZZ$ operation. Thus, if the qubit has not leaked, the state of the ancilla at the end of the gadget is $HZHX\ket{0}=\ket{0}$ and the ancilla $Z$-measurement outputs +1. If the data qubit has leaked before starting the circuit gadget, the two two-qubit gates are effectively deleted, leading to the action $HHX\ket{0}=\ket{1}$ on the ancilla. Thus, a $Z$-measurement on the ancilla deterministically measures whether the corresponding data qubit had leaked or not. The ancilla measurement outcomes are then used for post-selection, so that only shots that are not flagged as ``leaked" contribute to the reconstructed observables. These leakage gadgets are applied for the circuits containing 6 and 8 Trotter steps.

Between gate applications, the qubits can experience coherent and incoherent dephasing, as well as leakage.
To mitigate coherent dephasing, we use dynamical decoupling (DD) pulses.
In \texttt{Quantinuum System Model H2}, the compiler can be configured to automatically insert pairs of $X$ pulses opportunistically while qubits are shuttled through the gate zones while scheduling circuit operations.
This feature can be parameterized by specifying the longest idle time that we want the ions to experience during a given circuit execution; if the compiler determines that an ion is going to experience an idle time between operations that is longer than the threshold, it will try to schedule the $X$ pulses automatically. In our experiments, we set the threshold time to $0.03$ seconds, so that $X$-pulses are automatically inserted for idle windows that exceed the threshold time. 

\section{Convergence test of classical simulation}
Classical lattice model simulations are performed by time-dependent variational principle simulation with MPS ansatz \cite{tenpy, Haegeman2011TimeDependentVariationalPrinciple, Paeckel2019TimeevolutionMethodsMatrixproduct}. The convergence is performed against different bond dimensions $\chi$ and the size of time steps $dt$. We plot the time evolution of the string occupation probabilities $\mathcal{P}_{k=1}$, $\mathcal{P}_{k=1}$, and $\mathcal{P}_{k\ge3 }$ for different combinations of 
$\chi\in \{100, 200\}$ and $dt\in\{0.025, 0.05 \}$. The dynamics exhibit perfect overlap across all tested parameter regimes. This confirms that our chosen baseline parameters ($\chi=256$ and $dt=0.025$) are well within the converged regime and successfully capture the exact continuous-time quantum dynamics without being affected by finite-entanglement truncation or time-discretization errors.
\begin{figure}
    \centering
    \includegraphics[width=0.9\linewidth]{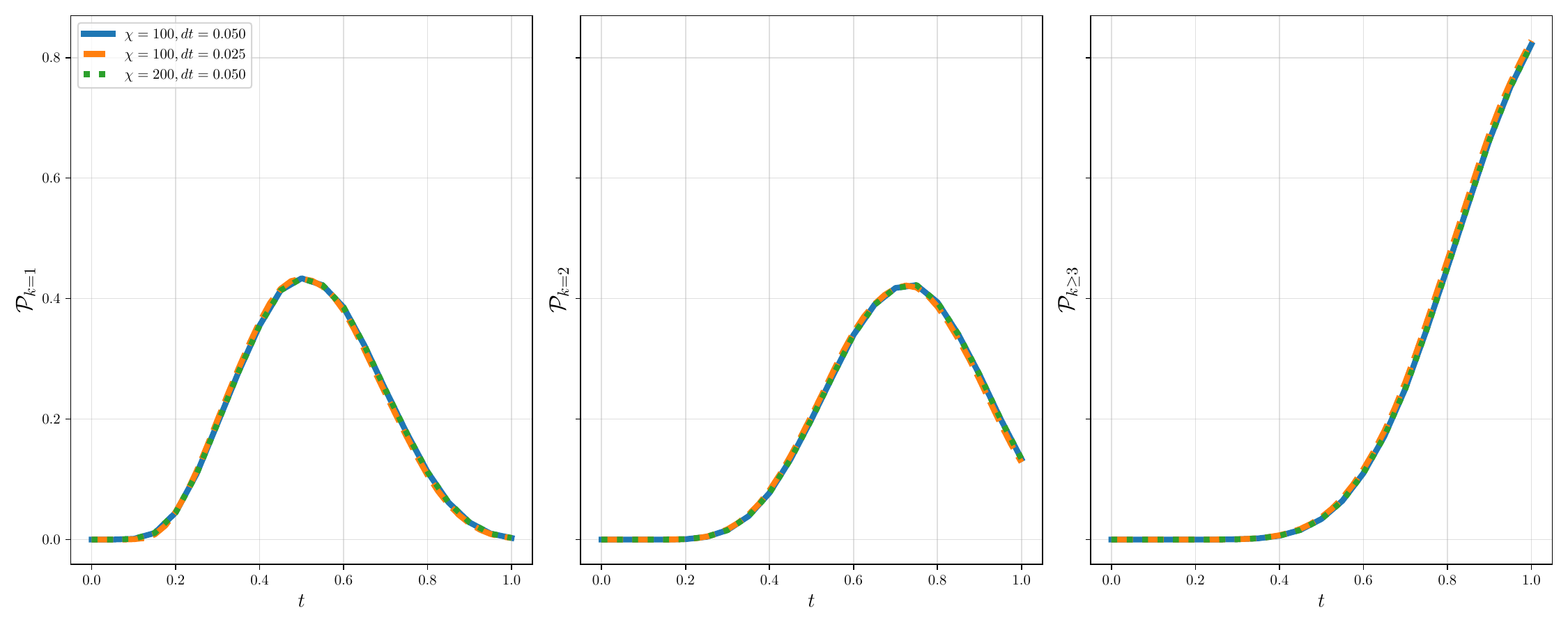}
    \caption{The convergence test against different bond dimensions and time steps. The parameters are chosen as $J_p=1$, $J_s=7$, $h_{E}=5$, $h_{M}=0.2$.}
    \label{fig:convergency}
\end{figure}

\section{The noise and noiseless emulation of small system size}

\begin{figure}[htbp]
    \centering
    \includegraphics[width=0.9\linewidth]{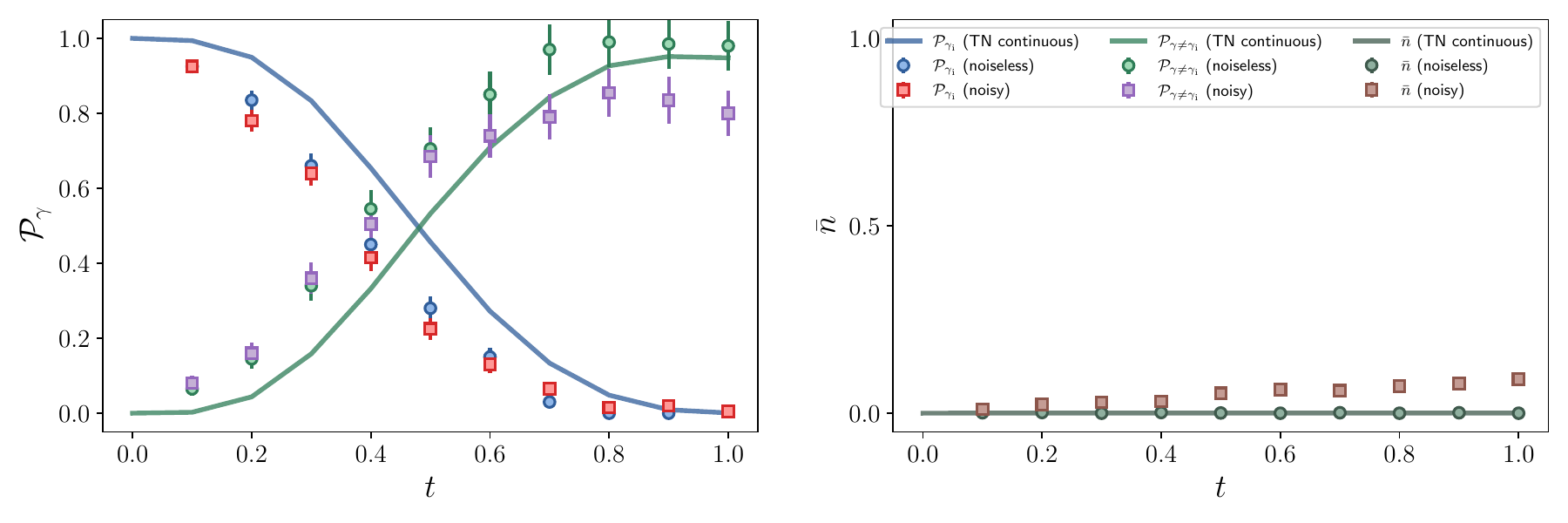}
    \caption{The comparison of state fidelity and minimal string probabilities population between continuous TN simulation, noiseless and noisy circuit emulation with $dt=0.1$ (left) and comparison between charge creation of TN and emulation results (right).}
    \label{fig:noiseless}
\end{figure}

To verify the validity of the Trotterized dynamics, we emulate noiseless Trotter circuits with 24 gauge links ($4\times4$ matter sites). We study the off-resonance quench dynamics of the minimal initial string with $J_p=1$, $J_s=7$, $h_E=5$, $h_M=0.2$. From Fig.~\ref{fig:noiseless}, one can observe that noiseless emulation shows a good match with continuous TN simulations except for a constant shift, as already argued in the discussion of the Trotter error. Compared to noiseless emulation, the noise model starts to show a noticeable effect after $t=0.5$ indicated by a small linear growth of charge creation and a degradation of string probabilities.

\section{A simple Quantum-Classical Hybrid Error Detection Circuit}

\begin{figure}[t!]
    \centering
    \includegraphics[width=0.8\linewidth]{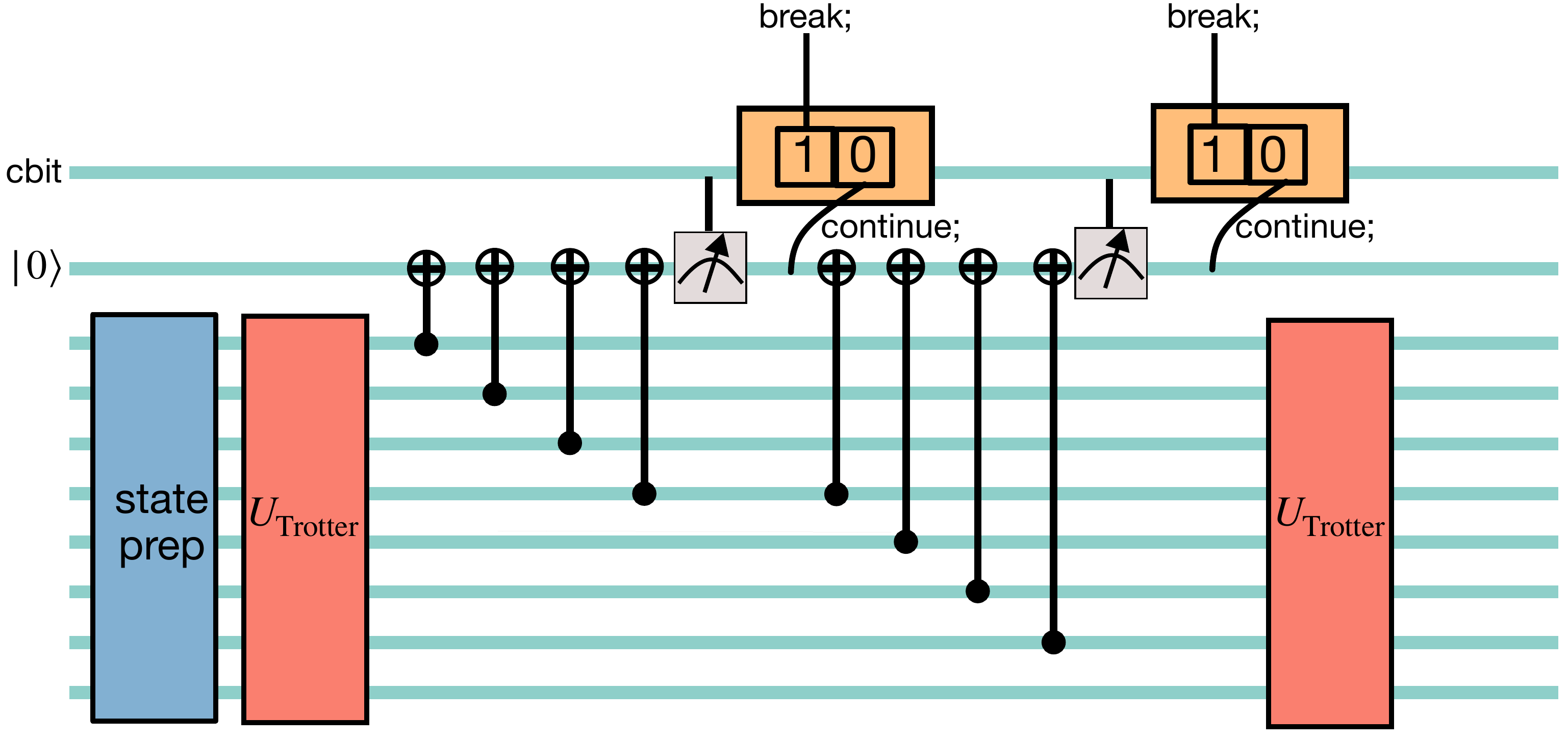}
    \caption{\textbf{Schematic of error detection circuit.} The parity check is composed of 4 CNOT gates with control on the vertex links and target on an ancilla qubit. The measurement is taken in $Z$-basis after every parity check of each vertex after the application of every Trotter box and the measurement outcome is mapped to a classical bit. If the measurement outcome is $0$ (vertex links are evenly occupied), continue this round, otherwise flag a possible one-qubit bit-flip error and terminate this shot. For simplicity, only qubits of two vertices are shown in the figure.}
    \label{fig:qc}
\end{figure}
\begin{figure}[t!]
    \centering
    \includegraphics[width=0.8\linewidth]{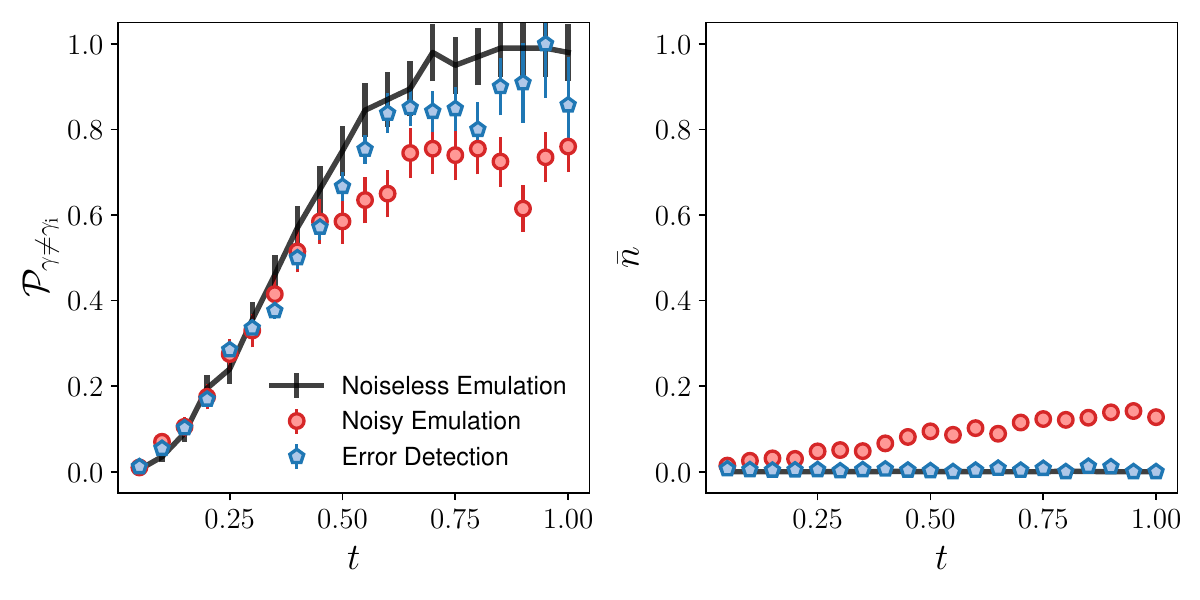}
    \caption{\textbf{Mitigation of hardware noise via error detection in the time evolution circuit}. The dynamics are simulated on a $4\times 4$ lattice using a Trotterized evolution up to time $t=1.0$ with time step $dt=0.05$ (Left). The probability of the system transitioning away from its initial string configuration, as a function of time. (Right) Time evolution of the average charge density $\bar{n}$, in both panels, the solid black line indicates the ideal, noiseless Trotter evolution. Raw data from the noisy hardware emulation (Quantinuum \texttt{H2-2E} emulator) are shown as red circles, displaying noticeable deviation from the ideal trajectory due to gate and measurement errors. The blue pentagons represent the results after applying an ancilla-based error detection and post-selection scheme. We choose the parameters as the same ones in the main text i.e., $J_s=1$, $J_p=7$, $h_{E}=5$, and $h_{M}=0.2$.}
    \label{fig:error-detection}
\end{figure}

Under the effective description with a sufficiently strong confining potential, the star operator act as a $\hat{Z}$ stabilizer, enforcing the gauge constraint that only even vertex parity is allowed. Based on this, we develop a Quantum-Classical hybrid error detection circuit, see Fig.~\ref{fig:qc}. The protocol can be demonstrated as follows: After each Trotter step, ancilla-assisted parity checks are performed on the vertices to verify whether the state remains within the zero charge subspace. If a vertex is detected in the -1 sector, the corresponding trajectory is discarded; otherwise, the circuit proceeds to the next Trotter step until the target stroboscopic time is reached. Such adaptive mid-circuit detection and conditional continuation can be implemented within the quantum-classical programming package \texttt{Guppy} \cite{koch2025guppypythonicquantumclassicalprogramming}. In principle, this method can detect the single-qubit bit-flip errors that creates matter at the vertices. We benchmark this protocol using emulation with the \texttt{H2-2} noise model in the deeply confined regime, see Fig.~\ref{fig:error-detection}. We find that post-selection on trajectories that remain in the confined effective subspace significantly improves the simulation fidelity, yielding dynamics that closely follow the noiseless evolution for both global string probabilities and local observables.

\end{document}